\documentclass[aps,showkeys,showpacs,superscriptaddress,twocolumn,floatfix]{revtex4}
\usepackage{graphicx}
\usepackage{amsmath}
\usepackage{amsfonts}
\usepackage{color}

\begin{document}

\title{Majority-vote model on triangular, honeycomb and Kagom\'e lattices}

\author{J. C. Santos}
\affiliation{
Universidade Federal do Piau\'{\i},
Departamento de F\'{\i}sica,\\
57072-970 Teresina, Piau\'{\i}, Brazil
}

\author{F. W. S. Lima}
\email{wel@ufpi.edu.br}
\affiliation{
Universidade Federal do Piau\'{\i},
Departamento de F\'{\i}sica,\\
57072-970 Teresina, Piau\'{\i}, Brazil
}

\author{K. Malarz}
\homepage{http://home.agh.edu.pl/malarz/}
\email{malarz@agh.edu.pl}
\affiliation{
AGH University of Science and Technology,
Faculty of Physics and Applied Computer Science,\\
al. Mickiewicza 30, PL-30059 Krak\'ow, Poland
}

\date{\today}

\begin{abstract}
On Archimedean lattices, the Ising model exhibits spontaneous ordering.
Three examples of these lattices of the majority-vote model with noise are considered and studied through extensive Monte Carlo simulations.
The order/disorder phase transition is observed in this system.
The calculated values of the critical noise parameter are $q_c=0.089(5)$, $q_c=0.078(3)$, and $q_c=0.114(2)$ for honeycomb, Kagom\'e and triangular lattices, respectively.
The critical exponents $\beta/\nu$, $\gamma/\nu$ and $1/\nu$ for this model are 
$0.15(5)$, $1.64(5)$, and $ 0.87(5)$; $0.14(3)$, $1.64(3)$, and $0.86(6)$; 
$0.12(4)$,
$1.59(5)$, and $1.08(6)$ for honeycomb, Kagom\'e and triangular lattices, respectively.
These results differs from the usual Ising model results and the majority-vote model on so-far studied regular lattices or complex networks.
The effective dimensionalities of the system $D_{\text{eff}}= 1.96(5)$ (honeycomb), $D_{\text{eff}} =1.92(4)$ (Kagom\'e), and $D_{\text{eff}}= 1.83(5)$ (triangular) for these networks are just compatible to the embedding dimension two. 
\end{abstract}

\pacs{
 05.10.Ln, 
 05.70.Fh, 
 64.60.Fr  
}

\keywords{Monte Carlo simulation, critical exponents, phase transition, non-equilibrium}

\maketitle

\section{Introduction}
 
The majority-vote model (MVM) \cite{MVM-SL} defined on two-dimensional regular lattices shows second-order phase transition with critical exponents $\beta$, $\gamma$, $\nu$ --- which characterize the system in the vicinity of the phase transition --- identical \cite{MVM-SL,MVM-regular,MVM-MFA} with those of equilibrium Ising model \cite{ising,critical}.

On the other hand MVM on the complex networks exhibit different behavior \cite{MVM-SW0,MVM-SW1,MVM-ERU,MVM-ERD,MVM-VD,MVM-ABU,MVM-ABD}.
Campos {\it et al.} investigated MVM on {\it undirected} small-world network \cite{MVM-SW0}.
This network was constructed using the square lattice (SL) by the rewiring procedure.
Campos {\it et al.} found that the critical exponents $\gamma/\nu$ and $\beta/\nu$ are different from those of the Ising model \cite{critical} and depend on the rewiring probability.
Luz and Lima studied MVM on {\it directed} small-world network \cite{MVM-SW1} constructed using the same process described by S\'anchez {\it et al.} \cite{sanchez}.
They also found that the critical exponents $\gamma/\nu$ and $\beta/\nu$ are different from these of the Ising model on square lattice, but contrary to results of Campos {\it et al.} \cite{MVM-SW0} for MVM the exponents do not depend on the rewiring probability.
Pereira {\it et al.} \cite{MVM-ERU} studied MVM on {\it undirected} Erd{\H o}s--R\'enyi's (ERU) classical random graphs \cite{ER}, and Lima {\it et al.} \cite{MVM-ERD} also studied this model on {\it directed} Erd{\H o}s--R\'enyi's (ERD) and their results obtained for critical exponents agree with the results of Pereira {\it et al.} \cite{MVM-ERU}, within the error bars.
Lima {\it et al.} \cite{MVM-VD} also studied this model on random Voronoy--Delaunay lattice \cite{VD} with periodic boundary conditions.
Lima also \cite{MVM-ABD} studied the MVM on {\it directed} Albert--Barab\'asi (ABD) network \cite{AB} and contrary to the Ising model on these networks \cite{alex}, the order/disorder phase transition {\em was} observed in this system.
However, the calculated $\beta/\nu$ and $\gamma/\nu$ exponents for MVM on ABD and ABU networks are different from those for the Ising model \cite{critical} and depend on the mean value of connectivity $\bar z$ of ABD and ABU network.
Lima and Malarz \cite{lima-malarz} studied the MVM on $(3,4,6,4)$ and $(3^4,6)$ Archimedean lattices (AL).
They remark that the critical exponents $\gamma/\nu$, $\beta/\nu$ and $1/\nu$ for MVM on $(3,4,6,4)$ AL are {\em different} from the Ising model \cite{critical} and {\em differ} from those for so-far studied regular two-dimensional lattices \cite{MVM-SL,MVM-regular}, but for $(3^4,6)$ AL, the critical exponents are much closer to those known analytically for SL Ising model.

The results presented in Refs. \cite{MVM-SW0,MVM-SW1,MVM-ERU,MVM-ERD,MVM-VD,MVM-ABU,MVM-ABD} show that the MVM on various complex topologies belongs to different universality classes.
Moreover, contrary for MVM on regular lattices \cite{MVM-SL,MVM-regular}, the obtained critical exponents are different from those of the equilibrium Ising model \cite{critical}.
Very recently, Yang and Kim \cite{yang2008} showed that also for $d$-dimensional hypercube lattices ($3\le d\le 6$) critical exponents for MVM differ from those for SL Ising model.
The same situation occurs on hyperbolic lattices \cite{wu2010}.

In this paper we study the MVM on three AL, namely on triangular $(3^6)$, honeycomb $(6^3)$, and Kagom\'e $(3,6,3,6)$ lattices.

The AL are vertex transitive graphs that can be embedded in a plane such that every face is a regular polygon.
The AL are labeled according to the sizes of faces incident to a given vertex.
The face sizes are sorted, starting from the face for which the list is the smallest in lexicographical order.
In this way, the triangular lattice gets the name $(3,3,3,3,3,3)$, abbreviated to $(3^6)$, honeycomb lattice is called $(6^3)$ and Kagom\'e lattice is $(3,6,3,6)$.
Critical properties of these lattices were investigated in terms of site percolation \cite{percolation} and Ising model \cite{zborek}.

Our main goal is to check the hypothesis of Grinstein {\it et al.} \cite{grinstein} --- i.e., that non-equilibrium stochastic spin systems with up-down symmetry fall in the universality class of the equilibrium Ising model --- for systems in-between ordinary, regular lattices (like SL \cite{MVM-SL}) and complex spin systems (like spins on ERU and ERD \cite{MVM-ERU,MVM-ERD} or ABU and ABD \cite{MVM-ABU,MVM-ABD}).

With extensive Monte Carlo simulation we show that MVM on $6^3$, $(3,6,3,6)$ and $3^6$ AL exhibits second-order phase transitions with effective dimensionality $D_{\text{eff}}\approx 1.96$, 1.92 and $1.83$ and has critical exponents that {\em do not} fall into universality class of the equilibrium Ising model.

\begin{figure*}[!hbt]
\begin{center}
\includegraphics[width=0.9\textwidth]{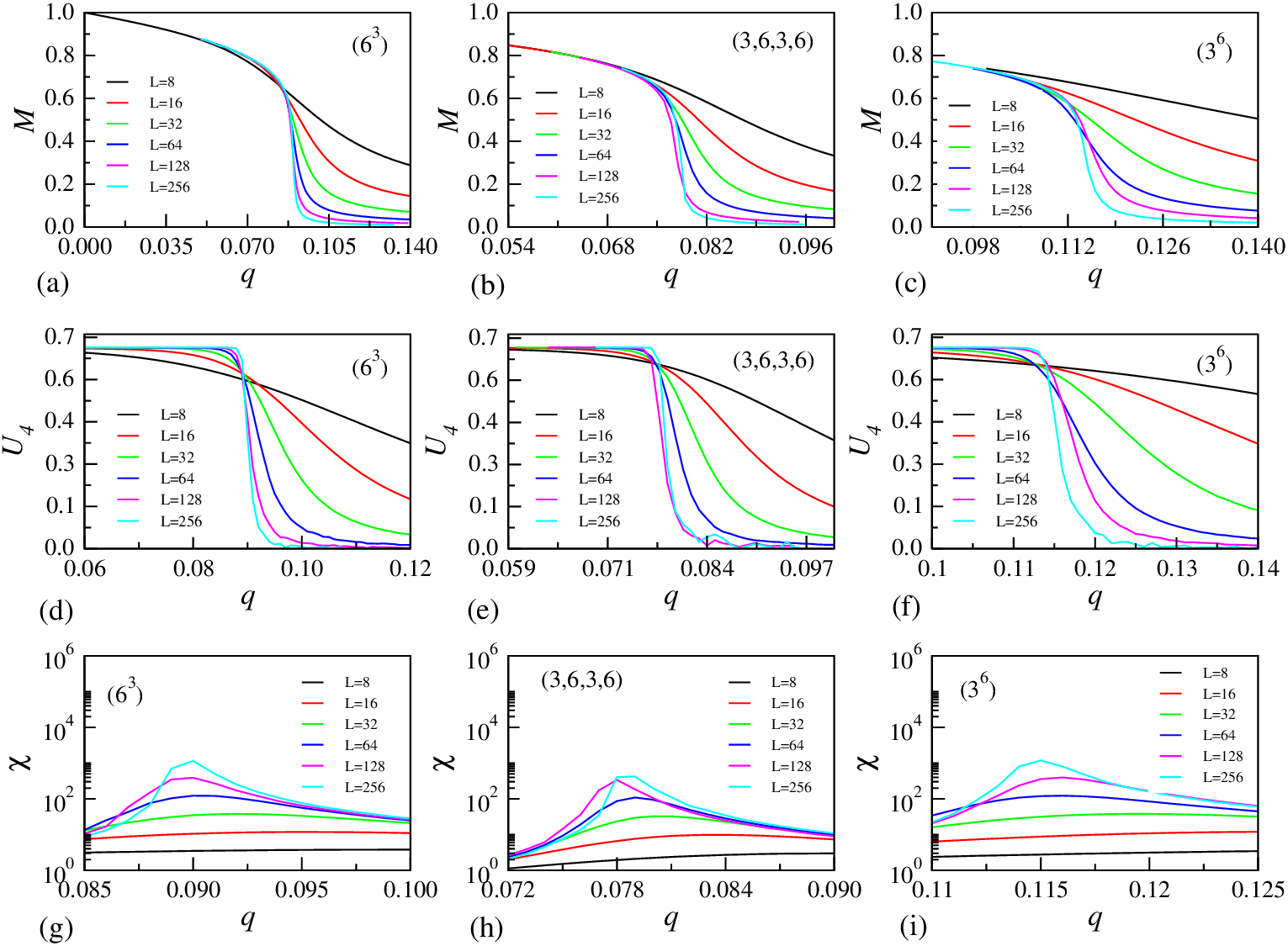}
\end{center}
\caption{\label{fig-M-U-C} (Color on-line). The magnetization $M$, Binder cumulant $U_4$, and susceptibility $\chi$ as a function of the noise parameter $q$, for $L=8$, $16$, $32$, $64$, $128$, and $256$ lattice sizes, and with $N=L^2$ sites for $6^3$ (first column), $3^6$ (second column) and $N=3L^2$ sites for $(3,6,3,6)$ AL (third column).}
\end{figure*}

\section{Model and simulation}

We consider the MVM \cite{MVM-SL} defined by a set of ``voters'' or spin variables $\sigma$ taking the values $+1$ or $-1$, situated on every node of the $6^3$, $(3,6,3,6)$ and $3^6$ AL with $N=L^2$ sites for $6^3$ and $3^6$, and $N=3L^2$ sites for $(3,6,3,6)$.
The evolution is governed by single spin-flip like dynamics with a probability $w_i$ of $i$-th spin to flip is given by
\begin{equation} 
w_i=\frac{1}{2}\left[ 1-(1-2q)\sigma_{i}\cdot\text{sign}\left(\sum_{j=1}^z\sigma_j\right)\right],
\end{equation}
and the sum runs over the number  $z=3$ (for $(3,6,3,6)$ and $(6^3)$ lattices) and $z=6$ (for $(3^6)$ lattice) of nearest neighbors of $i$-th spin.
The control parameter $0\le q\le 1$ plays the role of the temperature in equilibrium systems and measures the probability of aligning against the majority of neighbors.
It means, that a given spin $i$ adopts the majority sign of its neighbors with probability $(1-q)$ and the minority sign with probability $q$ \cite{MVM-SL,MVM-ERU,MVM-ERD,MVM-VD,MVM-ABD,MVM-ABU}.

To study the critical behavior of the model we define the variable $m\equiv\sum_{i=1}^{N}\sigma_{i}/N$.
In particular, we are interested in the magnetization $M$, susceptibility $\chi$ and the reduced fourth-order cumulant $U$
\begin{subequations}
\label{eq-def}
\begin{equation}
M(q)\equiv \langle|m|\rangle,
\end{equation}
\begin{equation}
\chi(q)\equiv N\left(\langle m^2\rangle-\langle m \rangle^2\right),
\end{equation}
\begin{equation}
U(q)\equiv 1-\dfrac{\langle m^{4}\rangle}{3\langle m^2 \rangle^2},
\end{equation}
\end{subequations}
where $\langle\cdots\rangle$ stands for a thermodynamics average.
The results are averaged over the $N_{\text{run}}$ independent simulations.

These quantities are functions of the noise parameter $q$ and obey the finite-size scaling relations
\begin{subequations}
\label{eq-scal}
\begin{equation}
\label{eq-scal-M}
M=L^{-\beta/\nu}f_m(x),
\end{equation}
\begin{equation}
\label{eq-scal-chi}
\chi=L^{\gamma/\nu}f_\chi(x),
\end{equation}
\begin{equation}
\label{eq-scal-dUdq}
\frac{dU}{dq}=L^{1/\nu}f_U(x),
\end{equation}
where $\nu$, $\beta$, and $\gamma$ are the usual critical 
exponents, $f_{m,\chi,U}(x)$ are the finite size scaling functions with
\begin{equation}
\label{eq-scal-x}
x=(q-q_c)L^{1/\nu}
\end{equation}
\end{subequations}
being the scaling variable.
Therefore, from the size dependence of $M$ and $\chi$ we obtained the exponents $\beta/\nu$ and $\gamma/\nu$, respectively.
The maximum value of susceptibility also scales as $L^{\gamma/\nu}$.
Moreover, the value of $q^*$ for which $\chi$ has a maximum is expected to scale with the system size as
\begin{equation}
\label{eq-q-max}
q^*=q_c+bL^{-1/\nu} \text{ with } b\approx 1.
\end{equation}
Therefore, the relations \eqref{eq-scal-dUdq} and \eqref{eq-q-max} may be used to get the exponent $1/\nu$.
We evaluate also the effective dimensionality, $D_{\text{eff}}$, from the hyper-scaling hypothesis
\begin{equation}
2\beta/\nu+\gamma/\nu=D_{\text{eff}}.
\label{eq-Deff}
\end{equation}

We performed Monte Carlo simulation on the $6^{3}$, $(3,6,3,6)$ and $3^{6}$ AL with various systems of size $N=64$, $256$, $1024$, $4096$, $16384$, and $65536$ for $6^{3}$ and $(3,6,3,6)$ AL and $N=192$, $768$, $3072$, $12288$, $49152$, and $196608$ for $3^{6}$.
It takes $2\times 10^5$ Monte Carlo steps (MCS) to make the system reach the steady state, and then the time averages are estimated over the next $2\times 10^5$ MCS.
One MCS is accomplished after all the $N$ spins are investigated whether they flip or not.
The results are averaged over $N_{\text{run}}$  $(20\le N_{\text{run}} \le 50)$ independent simulation runs for each lattice and for given set of parameters $(q,N)$.

\section{Results and Discussion}

\begin{figure}
\bigskip
\includegraphics[width=0.35\textwidth]{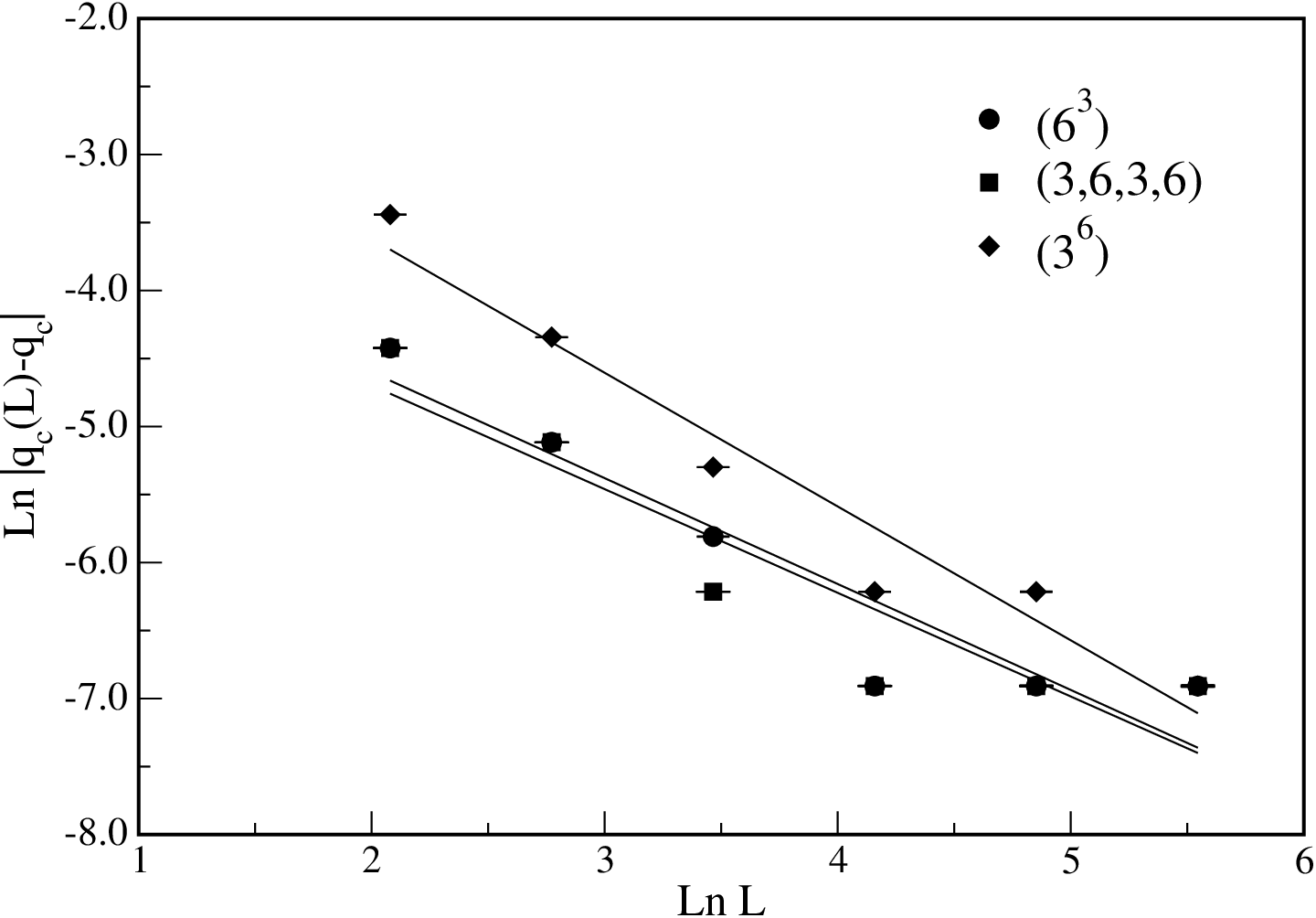}
\caption{\label{fig-eq-4} Plot $\ln|q_c(L)-q_c|$ versus the linear system size $L$ for $6^3$ (circles), $(3,6,3,6)$ (squares), $3^6$ (diamonds).}
\end{figure}

In Fig. \ref{fig-M-U-C} we show the dependence of the magnetization $M$, Binder cumulant $U_4$, and the susceptibility $\chi$ on the noise parameter $q$, obtained from simulations on $(6^3)$, $(3,6,3,6)$ and $(3^6)$ AL with $N$ ranging from $N=64$ to $196608$ sites.
The shape of $M(q)$, $U$, and $\chi$ curve, for a given value of $N$, suggests the presence of the second-order phase transition in the system.
The phase transition occurs at the value of the critical noise parameter $q_c$.
The critical noise parameter $q_c$ is estimated as the point where the curves for different system sizes $N$ intercept each other \cite{binder}.
Then, we obtain $q_c=0.089(5)$ and $U_4^*=0.578(3)$; $q_c=0.078(3)$ and $U_4^*=0.613(4)$; $q_c=0.114(5)$ and $U_4^*=0.601(5)$ for $6^3$, $(3,6,3,6)$ and $3^6$ AL, respectively.

\begin{figure}[!hbt]
\bigskip
\includegraphics[width=0.35\textwidth]{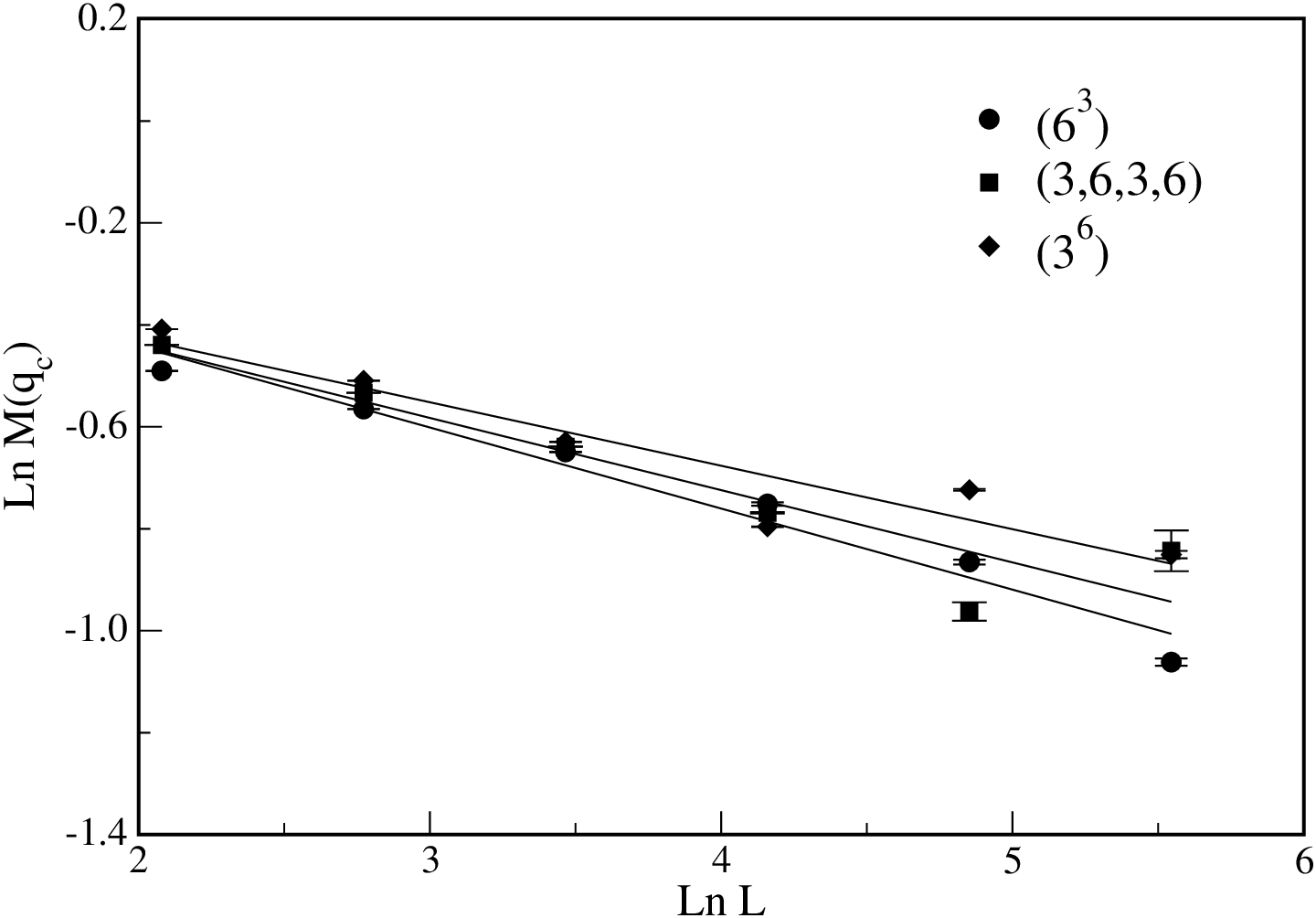}
\caption{\label{fig-M} Plot the dependence of the magnetization $M^*=M(q_c)$ vs. the linear system size $L$.}
\end{figure}

In Fig. \ref{fig-M} we plot the dependence of the magnetization $M^*=M(q_c)$ vs. the linear system size $L$.
The slopes of curves correspond to the exponent ratio $\beta/\nu$ according to Eq. \eqref{eq-scal-M}.
The obtained exponents are $\beta/\nu=0.15(5)$, $0.14(3)$, and $0.12(4)$, respectively for $6^3$, $(3,6,3,6)$ and $3^6$ AL.

\begin{figure}
\bigskip
\includegraphics[width=0.35\textwidth]{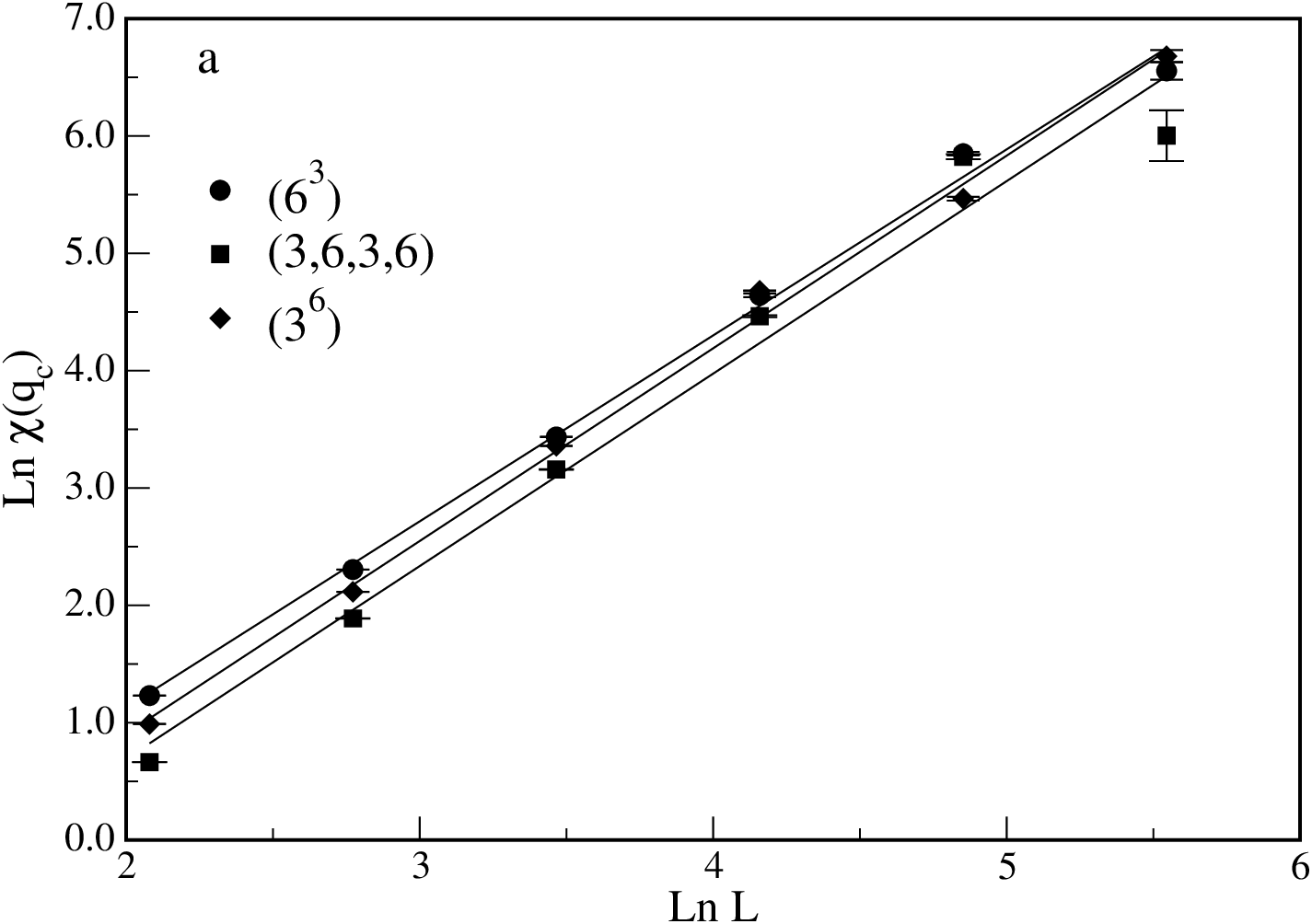}
\caption{\label{fig-chi-N1} Susceptibility at $q_c$ versus $L$ for  $(6^3)$, $(3,6,3,6)$ and $(3^6)$ AL.}
\end{figure}

\begin{figure}
\bigskip
\includegraphics[width=0.35\textwidth]{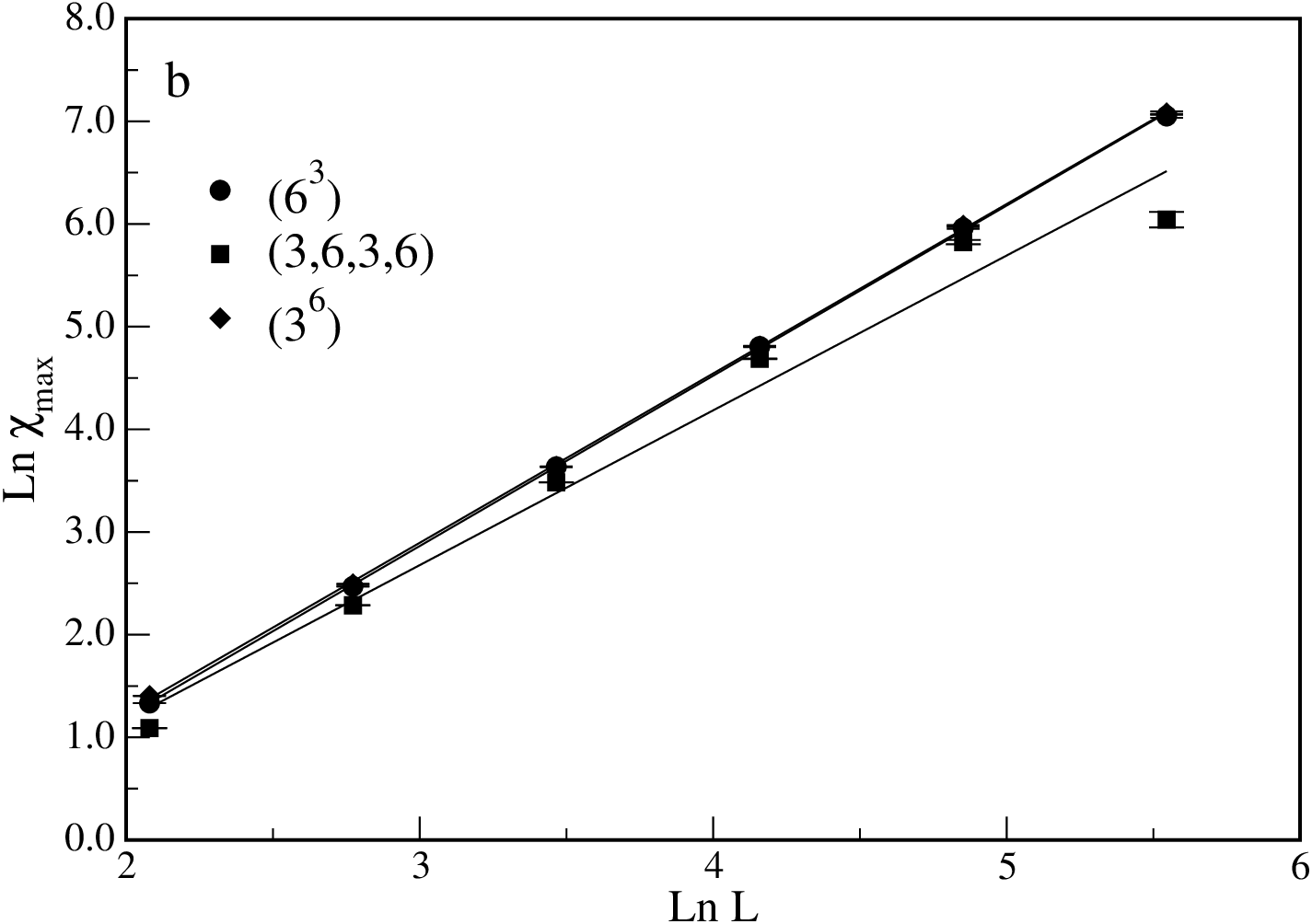}
\caption{\label{fig-chi-N2} Susceptibility at $q_{\chi_{max}}(N)$ versus $L$ for $6^3$, $(3,6,3,6)$ and $3^6$ AL.}
\end{figure}

The exponents ratio $\gamma/\nu$ at $q_{c}$ are obtained from the slopes of the straight lines with $\gamma/\nu=1.64(5)$ for $6^{3}$, $\gamma/\nu=1.64(3)$ for $(3,6,3,6)$, and $\gamma/\nu=1.59(5)$ for $3^{6}$, as presented in Fig. \ref{fig-chi-N1}. The exponents ratio $\gamma/\nu$ at $q_{\chi_{max}}(N)$ are $\gamma/\nu=1.66(8)$ for $6^{3}$, $\gamma/\nu=1.62(5)$ for $(3,6,3,6)$, and $\gamma/\nu=1.64(1)$ for $3^{6}$, as presented in Fig. \ref{fig-chi-N2}.

To obtain the critical exponent $1/\nu$, we used the scaling relation \eqref{eq-q-max}.
The calculated  values of the exponents $1/\nu$ are $1/\nu=0.87(5)$ for $6^{3}$ (circles),
 $1/\nu= 0.86(6)$ for $(3,6,3,6)$ (squares), and $1/\nu=1.08(6)$ for $3^{6}$ (diamonds) (see Fig. \ref{fig-eq-4}).
Eq. \eqref{eq-Deff} yields effective dimensionality of systems $D_{\text{eff}}= 1.96(5)$ for $6^{3}$, $D_{\text{eff}}=1.92(4)$ for $(3,6,3,6)$, and  $D_{\text{eff}}=1.83(5)$ for $3^{6}$.
The MVM on those three
 AL has the effective dimensionality close to two contrary to ER classical random graphs ($0.99\le D_{\text{eff}}\le 1.02$) \cite{MVM-ERU} or directed AB networks ($0.998\le D_{\text{eff}}\le 1.018$) \cite{MVM-ABU} with roughly the same nodes connectivity ($\bar z=3$) as for $(6^{3})$ and $(3,6,3,6)$, and ($\bar z=6$) $3^{6}$ AL.

The results of simulations
are collected in Tab. \ref{tab}.
 
\begin{table}
\caption{\label{tab} Critical parameter, exponents and effective dimension for MVM model on $6^3$, $(3,6,3,6)$ and $3^6$.
For completeness we cite data for SL Ising model as well.}
\begin{ruledtabular}
\begin{tabular}{r llllllll}
	& $6^3$ 
	& (3,6,3,6) 
	& $3^6$
	& SL Ising \\
\hline
$q_c$
	& 0.089(5) 
	& 0.078(2) 
	& 0.114(5)
	& ~ \\
$\beta/\nu$
	& 0.15(5) 
	& 0.14(3) 
	& 0.12(4)
	& 0.125 \\
$\gamma/\nu$\footnote{obtained using $\chi(N)$ at $q=q_c$}     
	& 1.64(5) 
	& 1.64(3) 
	& 1.59(5) 
	& 1.75 \\
$\gamma/\nu$\footnote{obtained using $\chi(N)$ at $q=q^*$}
	& 1.66(8) 
	& 1.62(5) 
	& 1.64(1)
	& 1.75 \\
$1/\nu$
	& 0.87(5) 
	& 0.86(6) 
	& 1.08(6)
	& 1 \\
$D_{\text{eff}}$\footnote{obtained using ratio $\gamma/\nu$ given by dependence $\chi(N)$ at $q=q_c$}
	& 1.96(5) 
	& 1.92(4) 
	& 1.83(5)
	& 2 \\
\end{tabular}
\end{ruledtabular}
\end{table}

\section{Conclusion}
 
We presented a very simple non-equilibrium MVM on $6^{3}$, $(3,6,3,6)$ and $3^{6}$ AL.
On these lattices, the MVM shows a second-order phase transition.
Our Monte Carlo simulations demonstrate that the effective dimensionality $D_{\text{eff}}$ is close to two, i.e. that hyper-scaling may be valid.

Finally, we remark that the critical exponents $\gamma/\nu$, $\beta/\nu$ and $1/\nu$ for MVM on {\em regular} $6^{3}$, $(3,6,3,6)$ and $3^{6}$ AL are {\em similar} to the MVM model on {\em regular} $(3,4,6,4)$ and $(3^4,6)$ \cite{lima-malarz} and are {\em different} from the Ising model \cite{critical} and {\em differ} from those for so-far studied regular lattices \cite{MVM-SL,MVM-regular} and for the {\it directed} and {\it undirected} ER random graphs \cite{MVM-ERU,MVM-ERD} and for the {\it directed} and {\it undirected} AB networks \cite{MVM-ABD,MVM-ABU}. 
However, in the latter cases \cite{MVM-ERU,MVM-ERD,MVM-ABU,MVM-ABD} the scaling relations \eqref{eq-scal} must involve the number of sites $N$ instead of linear system size $L$ as these networks in natural way do not posses such characteristic which allow for $N\propto L^d$ $(d\in\mathbb{Z})$ dependence \footnote{The linear dimension of such networks, i.e. its diameter --- defined as an average node-to-node distance --- grows usually logarithmically with the system size \cite{el-sw}.}.
For $(6^3)$, $(3,6,3,6)$ and $(3^6)$ AL some critical exponents are much closer to those known analytically for square lattice Ising model, i.e. $\beta=1/8=0.125$, $\gamma=7/4=1.75$ and $\nu=1$, but except for $\nu$ they differ for more than three numerically estimated uncertainties.

\begin{acknowledgments}
Authors are grateful to Dietrich Stauffer for stimulating discussions and for critical reading of the manuscript.
J.C.S. and F.W.S.L. acknowledge the support the system SGI Altix 1350 the computational park CENAPAD, UNICAMP-USP, SP-BRASIL and also the agency FAPEPI for the financial support.
K.M. acknowledges the machine time on SGI Altix 3700 in AGH University of Science and Technology, Academic Computer Center CYFRONET (grant No. MEiN/SGI3700/AGH/024/2006).
\end{acknowledgments}
 

\end{document}